\documentclass[pra,longbibliography,floatfix,superscriptaddress,reprint,]{revtex4-2}
\usepackage{datetime2}
\usepackage{amsmath}
\usepackage{amsfonts}
\usepackage{mathtools}
\usepackage{mathrsfs}
\usepackage{nicefrac}
\usepackage{xcolor}
\usepackage{float}
\usepackage{graphicx}
\usepackage[colorlinks=true,citecolor=blue]{hyperref}

\usepackage{physics}

\usepackage{booktabs}

\usepackage{dsfont}
\usepackage{bbm}

\usepackage[capitalize]{cleveref}

\usepackage[normalem]{ulem}
\usepackage{soul}

\def \addCQuIC {Center for Quantum Information and Control, University
  of New Mexico, Albuquerque, NM, 87131, USA}

\def \addCQuIC {Center for Quantum Information and Control, University of New Mexico, Albuquerque, 87131, NM, USA}
\def \addSandia {Sandia National Laboratories, Albuquerque, NM, 87185, USA}
\def \addPandAUNM {Department of Physics and Astronomy, University of New Mexico, Albuquerque, NM, 87106, USA}

\begin{document}

\title{Entangling quantum logic gates in neutral atoms via  the microwave-driven spin-flip blockade}

\author{Vikas Buchemmavari}
\email{bsdvikas@unm.edu}
\affiliation{\addCQuIC} \affiliation{\addPandAUNM}

\author{Sivaprasad Omanakuttan}
\affiliation{\addCQuIC} \affiliation{\addPandAUNM}

\author{Yuan-Yu Jau}
\affiliation{\addCQuIC} \affiliation{\addSandia} 

\author{Ivan Deutsch}
\email{ideutsch@unm.edu}
\affiliation{\addCQuIC} \affiliation{\addPandAUNM}

\date{2024/1/17}

\begin{abstract}

The Rydberg dipole-blockade has emerged as the standard mechanism to induce entanglement between neutral atom qubits. 
In these protocols, laser fields that couple qubit states to Rydberg states are modulated to implement entangling gates. 
Here we present an alternative protocol to implement entangling gates via Rydberg dressing and a microwave-field-driven spin-flip blockade [Y.-Y. Jau \textit{et al}, \href{https://doi.org/10.1038/nphys3487}{Nat. Phys. 12, 71 (2016)}]. 
We consider the specific example of qubits encoded in the clock states of cesium. 
An auxiliary hyperfine state is optically dressed so that it acquires partial Rydberg character. 
It thus acts as a proxy Rydberg state, with a nonlinear light-shift  that plays the role of blockade strength. 
A microwave-frequency field coupling a qubit state to this dressed auxiliary state can be modulated to implement entangling gates. 
Logic gate protocols designed for the optical regime can be imported to this microwave regime, for which experimental control methods are more robust. 
We show that unlike the strong dipole-blockade regime usually employed in Rydberg experiments, going to a moderate-spin-flip-blockade regime results in faster gates and smaller Rydberg decay. 
We study various regimes of operations that can yield high-fidelity two-qubit entangling gates and characterize their analytical behavior. 
In addition to the inherent robustness of microwave control, we can design these gates to be more robust to laser amplitude and frequency noises at the cost of a small increase in Rydberg decay.
\end{abstract}

\maketitle
\section{Introduction}\label{sec:1:Intro}




Arrays of optically trapped atoms have emerged as a powerful platform for quantum information processing (QIP)~\cite{Brennen1999, Deutsch2000, Jaksch_gate_2000, Saffman_review_2016_Rydberg,Browaeys_review_2020_Quantum}. 
This architecture has a number of unique capabilities including the ability to operate on arrays with hundreds of atoms~\cite{Ebadi_Lukin_Nature_2021_256_atom,Scholl_Browaeys_Nature_2021_Q_simulation}, the ability to reconfigure the geometry through atom transport~\cite{Barredo_Browaeys_Atom_assembly,Endres_Lukin_Atom_assembly,Lukin_Nature_2022}, and possibilities of multiple atomic species and mechanisms with which to encode, manipulate, and measure quantum information~\cite{Saffman_Nature_2022,Levine_Pichler_gate,Madjarov_Endres_2020_Sr,Lis_Senoo_Kaufman_Yb_mid_circuit_2023,Ma_Thompson_Yb_erasure_gate,Singh_Bernien_2022_dual_array,Siva_Qudit_entangler_2023,Atom_computing_2022}.  Applications include quantum metrology~\cite{Schine_Kaufman_Bell_state_Martin,Kaubruegger_Zoller_metrology,Bornet_Browaeys_2023_Scalable_sqeezing,Hines_Schleier-Smith_2023_Squeezing_Rydberg_dressing, Eckner_Kaufman_2023_Squeezing_clock}, simulation of many-body physics~\cite{Scholl_Browaeys_Nature_2021_Q_simulation,Ebadi_Lukin_Nature_2021_256_atom,Bornet_Browaeys_2023_Scalable_sqeezing,Browaeys_many_body_review_2020_Nature,Bornet_Browaeys_2023_Scalable_sqeezing}, optimization of combinatoric problems~\cite{Kim_Ahn_2022_Nature_Rydberg_wores_MIS,Ebadi_Lukin_2022_MIS_optimization,Optimization_Pichler_2022_PRXQ}, and  universal quantum computing with potential paths to  fault-tolerant error correction~\cite{Wu_Puri_Thompson_2022_Nature_erasure,Cong_Lukin_QEC_Rydberg_PRX_2022,Lukin_Nature_2022}.   The development of neutral-atom quantum computing has paralleled the development of atomic clocks,  both in the traditional alkali-metal atoms (cesium and rubidium) that define microwave frequency standards  and in alkaline-earth-like atoms (strontium and ytterbium) that define optical frequency standards~\cite{Wineland_Nobel_Lecture_2013,Haroche_Nobel_Lecture_2013,Bloom_Ye_2014_Nature_clock}.  While the latter provide a path to ultra precise clocks,  manipulating qubits  at microwave frequencies offers potential advantages for coherent control required for QIP and will be the focus of this work.


The standard entangling gates for neutral-atom QIP are based on Rydberg dipole blockade, arising from strong electric dipole-dipole interaction, as originally proposed in~\cite{Jaksch_gate_2000}. 
A plethora of gate protocols and variations have since been proposed, including those in~\cite{Gate_EIT_Muller_Zoller_PRL_2009,Gate_CkNOT_Molmer_Saffman_QIP_2011,Gate_Blockade_error_free_Shi_PRApplied_2017,Gate_dark_state_Molmer_Saffman_PRA_2017,Beterov_Saffman_Ryabtsev_Forster_stark_2016_PRA,Saffman_Sanders_symmetric_adiabatic_PRA_2020,Mitra_Martin_gate}. 
The leading protocol is currently the Levine-Pichler gate, which is based on the geometric phases accumulated through Rabi oscillations between qubit states and Rydberg states in closed-loop trajectories on the Bloch sphere~\cite{Levine_Pichler_gate,Jandura_Pupillo_Optimal_control, Pagano_strontium_dCRAB}. 
While the fidelity is fundamentally limited by the lifetime of Rydberg states,  the current gate infidelities are dominated by other noise sources including inhomogeneities due to atomic thermal motion, laser noise, and imperfect shielding of stray electric fields.
 Quantum optimal control and dynamical decoupling provide powerful methods to mitigate some of these imperfections,~\cite{ jandura_Pupillo_robust,fromonteil_Bluvstein_Pichler_robust,Mohan_Kokkelmans_Robust_control_PRR_2023}. 
With these tools in hand, recent experiments now demonstrate gate fidelities of up to 0.995~\cite{Evered_Lukin_2023_High_fidelity,Ma_Thompson_Yb_erasure_gate}.

While impressive progress has been made, current experiments are challenging, and somewhat constrained by the conditions required to implement optimal control at optical frequencies.  As such, it is useful to consider additional method that can be more easily generalized, particularly by employing the more robust control methods in the microwave or radio-frequency regime. To achieve this, we consider an alternative protocol that  employs Rydberg dressing~\cite{Johnson_Rolston_dressing, Pohl2010} on one of the qubit states. Due to the strong  dipole-dipole interaction when both atoms are in the Rydberg state, the resulting light shift of two dressed atoms is different from two independently dressed atoms.  The difference in the light shift energies $J$  provides an entangling mechanism~\cite{Keating_gate,Mitra_Martin_gate,Schine_Kaufman_Bell_state_Martin}. Of particular interest is the  phenomenon of the spin-flip blockade for qubits encoded in the clock states of alkali-metal atoms~\cite{Jau_Nature_2016}.  In the presence of Rydberg dressing, a microwave photon can flip the qubit spin from the lower  to the upper hyperfine state, but the spin of two qubits is blockaded when $J$ is sufficiently large.  Like its optical Rydberg-blockade counterpart\cite{Madjarov_Endres_2020_Sr}, the microwave spin-flip blockade provides a direct route to creating Bell states~\cite{Jau_Nature_2016}.

In this paper, we extend the use of the spin-flip blockade from Bell-state preparation to full two-qubit logic gates.   A new ingredient is the introduction of an auxiliary dressed hyperfine level that plays the role of the Rydberg state but for microwave-driven transitions. Importantly, this allows us to map the dipole-blockade physics implemented in the optical regime to the spin-flip blockade in the microwave regime. Hence, any gate protocols that can be implemented in the optical regime, using an optical/UV field coupling to the Rydberg state, can be implemented in the microwave regime. While this does not improve the fundamental fidelity limited by Rydberg state lifetime, it opens the door to more robust implementations as one can employ the mature tools of microwave control, and reduce noise from residual Doppler shifts and  other inhomogeneities.   These, by far, are the main sources of infidelity in current implementations.


The remainder of this paper is organized as follows. 
In Sec.~\ref{sec:Mapping_from_optical_to_microwave_control} we show how to map the physics of the Rydberg blockade to the spin-flip blockade using dressed states, which forms the basis of our gate protocols. 
In Sec.~\ref{sec:2:Entangling_gate} we consider specific gate protocols and use of quantum control to optimize performance.
In Sec.~\ref{sec:Robust_control} we consider the most important sources of noise due to thermal fluctuations in the atomic motion, detuning errors, and other inhomogeneities, and show how robust optimal control can be used to safeguard against these imperfections.  
Section~\ref{sec:summary_and_outlook} summarizes the results and presents an outlook for future work.

\section{Mapping from optical to microwave control}
\label{sec:Mapping_from_optical_to_microwave_control}

\begin{figure*}
  \makebox[\textwidth]{
    \includegraphics[width=0.85\textwidth]{  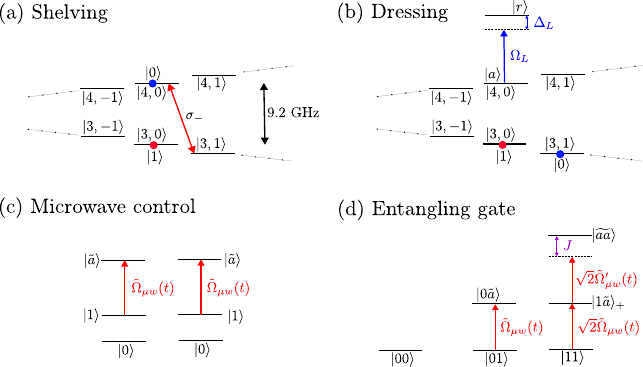}}
  \caption{Level diagram for implementing entangling gates in the microwave regime. (a) The hyperfine states of the cesium electronic ground state, labeled $\ket{F,M_F}$, are shown in the presence of a bias magnetic field, near the clock states with $M_F=0$ in which a qubit is encoded.  Before implementing an entangling gate, the logical state $\ket{0}$ is shelved in $\ket{3,1}$ through a microwave resonance. (b) $\ket{4,0}$ is now an auxiliary state $|a\rangle$, dressed with a Rydberg laser (Rabi frequency $\Omega_\mathrm{L}$, detuning $\Delta_\mathrm{L}$). (c) The phase of a microwave/Raman field of correct frequency and polarization couples $|1\rangle$ and dressed state $|\tilde{a}\rangle$, leaving $\ket{0}$ untouched, is modulated to implement the gate.  The microwave here is shown tuned to resonance with the dressed state and the Rabi frequency, $\tilde{\Omega}_{\mu w}(t)$, is modified from its bare value due to the admixture of $\ket{r}$ in $|\tilde{a}\rangle$. (d) Two-atom symmetric basis showing the coupling of the logical basis to the Rydberg-dressed states. The state $|01\rangle$($\ket{10}$) couples to the dressed state $|0\tilde{a}\rangle$($\ket{\tilde{a}0}$) with modified Rabi frequency $\tilde{\Omega}_{\mu w}$ . The two-atom state $|11\rangle$ couples to the bright state $|\tilde{a}1\rangle_+$ and it to $|\widetilde{aa}\rangle$ with modified Rabi frequencies $\sqrt{2}\tilde{\Omega}_{\mu w},\sqrt{2}\tilde{\Omega}'_{\mu w}$. The state $|\widetilde{aa}\rangle$ has a nonlinear light shift that we call the entangling energy, $J$, which determines the strength of the atom-atom coupling.
  }\label{fig:Protocol}
\end{figure*}

\begin{figure}
    \centering
    \includegraphics[width=1.1\columnwidth]{  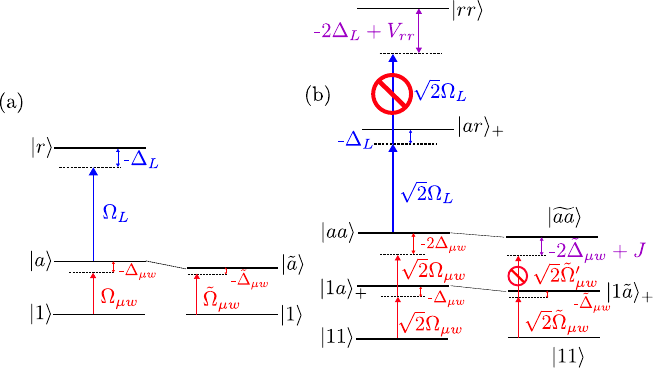}
    \caption{Dressing scheme to map the optical Rydberg blockade to the microwave spin-flip blockade. (a) Single atom level structure.  Electronic ground states $\ket{a}$ and $\ket{1}$ (e.g., hyperfine clock states) can be coupled by a microwave/Raman field with Rabi frequency $\Omega_{\mu w}$ and detuning $\Delta_{\mu w}$.  The ``auxiliary" state $\ket{a}$ is dressed through optical coupling to Rydberg state $\ket{r}$ with Rabi frequency $\Omega_\mathrm{L}$ and detuning $\Delta_L$, leading to a one-atom light shift $E_{\tilde{a}}$.  The admixture of Rydberg character modifies the microwave coupling Rabi frequency $\tilde{\Omega}_{\mu w} = \langle a | \tilde{a} \rangle {\Omega}_{\mu w}$ and detuning $-\tilde{\Delta}_{\mu w} =-\Delta_{\mu w} + E_{\tilde{a}}$.  (b) Level diagram in the two-atom symmetric subspace, including atom-atom interactions.  In the symmetric state $\ket{1a}_+ =(\ket{1a}+\ket{a1})/\sqrt{2}$, only one atom can be optically dressed through coupling to $\ket{1r}_+$ with Rabi frequency $\Omega_\mathrm{L}$, leading to dressed state $\ket{1\tilde{a}}$.  The state $\ket{aa}$, in which two atoms can be optically excited, is coupled to $\ket{ar}_+$ with Rabi frequency $\sqrt{2}\Omega_L$, but this state is blockaded from excitation to $\ket{rr}$ by the Van der Waals energy $V_{rr}$.  The result is a two-atom light shift $E_{\widetilde{aa}}= 2E_{\tilde{a}} + J$, leading to the dressed ground-state $\ket{\widetilde{aa}}$  shown on the right.  The state $\ket{11}$ is coupled to $\ket{1\tilde{a}}_+$ by microwave/Raman photons, but the state $\ket{1\tilde{a}}_+$ is blockaded from excitation to $\ket{\widetilde{aa}}$ due to the nonlinear light shift $J$ (the spin-flip blockade).  The level diagram for the optically-coupled triplet $\{\ket{aa}, \ket{ar}_+, \ket{rr}\}$ maps directly to the microwave-coupled dressed triplet $\{\ket{11}, \ket{1\tilde{a}}_+,\ket{\widetilde{aa}}\}$ level diagram.
    }   
    \label{fig:Dressing_mapping}
\end{figure}

\begin{figure}
    \centering
    \includegraphics[width=1.1\columnwidth]{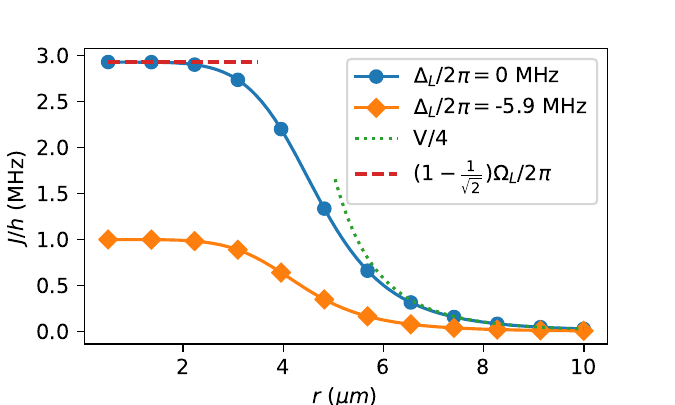}
    \caption{The entangling energy, $J$, as a function of the interataomic distance $r$. For small interatomic separations, $V_{rr}\gg\Omega_\mathrm{L}$, resulting in  strong Rydberg blockade. 
    For resonant dressing (blue dots), this results in the plateau of $J_{\mathrm{max}}=(1-\frac{1}{\sqrt{2}})\Omega_\mathrm{L}$. 
  As the distance increases, the interaction energy falls off rapidly ($V_{rr}=C_6/r^{6}$) and results in a steep drop-off in the value of $J$. 
  At large distances, when $V_{rr}\ll\Omega_L$, $J=V_{rr}/4$, and hence falls off as $r^{-6}$\cite{Mitra_Omanakuttan_adiabatic_gates}.
  A similar behaviour is seen for non-resonant dressing (orange diamonds), but with a smaller plateau value of $J$. 
  We show the behavior of $J(r)$ for $\Delta_L/2\pi =-5.9\mathrm{MHz}$, where the plateau $J$ value is equal to $\Omega_{\mu w}=1\;\mathrm{MHz}$. 
  The examples shown are for the chosen Rydberg state $^{133}$Cs $64\mathrm{P}_{3/2}, m_\mathrm{J}=3/2$, for which $C_6=677\;\mathrm{GHz}/\mu m^6$\cite{ARC} and $\Omega_\mathrm{L}/2\pi=10\;\mathrm{MHz}$.
    }   
    \label{fig:J_vs_r}
\end{figure}

To begin we describe how coherent control associated with the Rydberg-blockade in the optical regime can be mapped to the microwave regime via Rydberg-dressing and the spin-flip blockade. 
For concreteness, we consider $^{133}\mathrm{Cs}$ where qubits are stored in the hyperfine levels of the electronic ground state $6S_{1/2}$, though this easily generalizes to other species for both alkali and alkaline-earth-like atoms. 
We take the logical states to be the clock states $\ket{0}=\ket{F=4,m_F=0}$, $\ket{1}=\ket{F=3,m_F=0}$. 

In the presence of a bias magnetic field, when a two-qubit gate is to be performed, the population in the $\ket{0}$-state of each atom can be shelved in the $\ket{F=3,m_F=1}$ state, where it is effectively removed from further interactions, as in Fig.~\ref{fig:Protocol}(a).
The now-empty state $\ket{F=4,m_F=0}$ is then repurposed as the auxiliary state $\ket{a}$, which is coupled to a Rydberg state $\ket{r}$ via a near-resonance optical/UV laser field, as shown in Fig.~\ref{fig:Protocol}(b).
We choose our shelving protocol and auxiliary state $\ket{a}$ such that when we dress the auxiliary state, the populations in the computational states remain undisturbed, and the transition between states $\ket{1}$ and $\ket{\tilde{a}}$ is controllable. This choice would vary depending on the qubit encoding and atomic species. For example, for a qubit encoded in the nuclear spin of the ground $^1\mathrm{S}_0$ manifold of  $^{171}\mathrm{Yb}$, one can simply dress one of the metastable $^3\mathrm{P}_0$ clock states as the auxiliary state, without the need for shelving in a clock-laser-driven entangling gate. 

For a single atom addressed by the dressing laser with detuning $\Delta_\mathrm{L}$ and  Rabi frequency $\Omega_\mathrm{L}$, this results in a dressed Autler-Townes doublet, as shown in Fig.~\ref{fig:Dressing_mapping}(a), 
\begin{equation} \label{eqn: 1-atom dressing}
\begin{aligned}
    \ket{\tilde{a}} &= \cos\frac{\theta}{2}\ket{a}+\sin\frac{\theta}{2}\ket{r}, \\
    \ket{\tilde{r}} &= \cos\frac{\theta}{2}\ket{r}-\sin\frac{\theta}{2}\ket{a}, 
\end{aligned} 
\end{equation}
where $\tan\theta  = -\Omega_\mathrm{L}/\Delta_\mathrm{L}$.
These dressed states are light-shifted by energies  $-\Delta_\mathrm{L}/2 \pm \sqrt{\Omega_\mathrm{L}^2 +\Delta_\mathrm{L}^2}/2$ (here and throughout $\hbar=1$).  Due to the admixture, the dressed state, $\ket{\tilde{a}}$, then plays the role of the Rydberg-state for entangling the spin qubits.  The states  $\ket{1}$ and  $\ket{\tilde{a}}$ can be coupled by a microwave/Raman field with an effective Rabi frequency  $\tilde{\Omega}_{\mu w}=\cos(\theta/2)\Omega_{\mu w}$, where $\Omega_{\mu w}$ is the Rabi frequency coupling $\ket{1}$ to the bare state $\ket{a}$. When driving this transition resonantly, if $\sqrt{\Omega_\mathrm{L}^2+\Delta_\mathrm{L}^2}\gg \tilde{\Omega}_{\mu w}$, we can neglect the coupling from $\ket{1}$ to $|\tilde{r}\rangle$. 

To entangle two qubits we consider symmetrically dressing the atoms by the same uniform laser and microwave/Raman fields.  Interactions between the qubits only occur when both atoms are initially in the $\ket{1}$ state and are then excited by microwaves.  To describe the entangling interaction, we need only to consider the symmetric subspace, $\ket{11}$, $\ket{1a}_+$, and $\ket{aa}$;  here and throughout $\ket{xy}_\pm \equiv (\ket{xy}\pm\ket{yx})/\sqrt{2}$.   The $\ket{1a}_+$ state is coupled to $\ket{1r}_+$ with Rabi frequency $\Omega_\mathrm{L}$, yielding  the same single atom dressed states as Eq. (\ref{eqn: 1-atom dressing}), denoted $\ket{1\tilde{a}}_+$ and $\ket{1\tilde{r}}_+$.  The state $\ket{aa}$ is coupled to $\ket{ar}_+$  with Rabi frequency $\sqrt{2} \Omega_\mathrm{L}$.  For conditions well-approximated by a perfect Rydberg blockade, the state $\ket{rr}$ can be ignored and the resulting dressed states are denoted $\ket{\widetilde{aa}}$,  $\ket{\widetilde{ar}}_+$. This doublet  is split by $\sqrt{2\Omega_\mathrm{L}^2+\Delta_\mathrm{L}^2}$.  The nonlinear behavior of the light shift, due to the Rydberg blockage, implies that the energy of the two interacting atoms is not equal to the sum of the energies of the non-interacting atoms. The difference is the  ``entangling energy,'' defined as $J=E_{\widetilde{aa}}-2E_{1\tilde{a}_+} = \frac{1}{2}[\Delta_\mathrm{L} \pm(\sqrt{2\Omega_\mathrm{L}^2 + \Delta_\mathrm{L}^2} - 2 \sqrt{\Omega_\mathrm{L}^2 + \Delta_\mathrm{L}^2})]$ (see Fig.~\ref{fig:J_vs_r}). The $\pm$ sign corresponds to two branches of the light shift, adiabatically connected to the red or blue side of resonance~\cite{Mitra_Martin_gate}.

Critically, in the presence of dressing, when driving the microwave transition $\ket{11}\rightarrow\ket{1\tilde{a}}_+$ with Rabi frequency $\tilde{\Omega}_{\mu w}$, the transition $\ket{1\tilde{a}}_+\rightarrow \ket{\widetilde{aa}}$ will be blockaded when $J\gg \tilde{\Omega}_{\mu w}$.  This is the spin-flip blockade, as first observed in~\cite{Jau_Nature_2016}.  As seen in Fig.~\ref{fig:Dressing_mapping}, the level structure of the dressed states mimics that of the optically driven ground-Rydberg system with the dressed auxiliary state as a proxy for the Rydberg state, and the spin-flip-blockade replacing the Rydberg blockade. This allows us to map all gate protocols from the optical regime to the microwave regime.

While the essential physics is well understood assuming a perfect Rydberg blockade, this may not be the optimal operating point given the fundamental limit on the gate fidelity determined by the decay rate of Rydberg state $\Gamma_r$.  More generally  when two interacting atoms participate in dressing, including some admixture of doubly-excited Rydberg states, the result is the dressed triplet,
\begin{equation}
\begin{aligned}
    \ket{\widetilde{aa}} &= \alpha\ket{aa} +\beta\ket{ar}_++\gamma\ket{rr}, \\
    \ket{\widetilde{ar}}_+ &= \alpha_2\ket{ar}_+ +\beta_2\ket{aa}+\gamma_2\ket{rr}, \\
    \ket{\widetilde{rr}} &= \alpha_3\ket{rr} +\beta_3\ket{aa}+\gamma_3\ket{ar}_+,
\end{aligned}
\end{equation}
with dressed energies $E_{\widetilde{aa}}$, $E_{\widetilde{ar}_+}$ and $E_{\widetilde{rr}}$ respectively. 
In this case the Rabi frequency for the coupling $\ket{11}\leftrightarrow\ket{1\tilde{a}}_+$ is given by $\sqrt{2}\tilde{\Omega}_{\mu w}= \sqrt{2}(\cos\frac{\theta}{2})\Omega_{\mu w}$ and for $\ket{1\tilde{a}}_+\leftrightarrow\ket{\widetilde{aa}}$,  $\sqrt{2}\tilde{\Omega}'_{\mu w}= \sqrt{2}(\alpha \cos\frac{\theta}{2}+\beta\sin\frac{\theta}{2}/\sqrt{2})\Omega_{\mu w}$.  
Again, for appropriate microwave detuning (equal to the dressed light-shift on the state $\ket{a}$ $E_{\tilde{a}}$) we can neglect coupling of  $\ket{1\tilde{a}}_+$ to $|\widetilde{ar}\rangle_+$ and $\ket{\widetilde{rr}}$.   
Note, since the  states $\ket{\tilde{a}},\ket{\widetilde{aa}}$ have only partial Rydberg character, their effective decay rates will be proportional to the admixture of Rydberg state in the dressing,
\begin{equation}
    \begin{aligned}
    \Gamma_{\tilde{a}}&=\sin^2\frac{\theta}{2}\Gamma_r \\
    \Gamma_{\widetilde{aa}}&=(|\beta|^2+2|\gamma|^2)\Gamma_r
\end{aligned}
\end{equation}
where we assumed that the state $\ket{rr}$ decays twice as fast as state $\ket{r}$. 

In summary, through Rydberg dressing the optical physics of the dipole blockade is mapped to the microwave spin-flip blockade, with $\ket{\tilde{a}}$ playing the role of $\ket{r}$  according to mapping:
\begin{equation}
    \begin{aligned}
    \Omega_\mathrm{L}&\rightarrow \tilde{\Omega}_{\mu w},\tilde{\Omega}'_{\mu w}\\
    V_{rr}&\rightarrow J \\
    \Gamma_r&\rightarrow\Gamma_{\tilde{a}} \\
    \Gamma_{rr}&\rightarrow \Gamma_{\widetilde{aa}}
\end{aligned}
\end{equation}

 One key difference is the ability to control the entangling energy, $J$, by controlling the laser parameters $\Omega_\mathrm{L}, \Delta_\mathrm{L}$ and the well defined nature of the state $\ket{\widetilde{aa}}$ (see Fig.~\ref{fig:J_vs_r}).  This is in contrast to the optical regime where the doubly-excited Rydberg spectrum can be complex and $V_{rr}$ is strongly dependent on the interatomic distance.  We will use this to our advantage below by operating in the moderate-blockade regime where $J\approx\Omega_{\mu w}$.
 
\section{Entangling gate protocol}\label{sec:2:Entangling_gate}

In this section we use the optical-to-microwave mapping described above as a vehicle for implementing two-qubit entangling gates. Previous work was based on adiabatic dressing~\cite{Keating_Robust_computation, Mitra_Martin_gate,Keating_gate}, which is intrinsically robust to certain forms of noise. Here we demonstrate the versatility of Rydberg dressing and microwave control in its application of the most successful gate to date, the Levine-Pichler (LP) gate~\cite{Levine_Pichler_gate}. The LP-gate is also well suited to optimal control, as was theoretically studied in~\cite{Jandura_Pupillo_Optimal_control,Pagano_strontium_dCRAB} and recently demonstrated in~\cite{Evered_Lukin_2023_High_fidelity,Ma_Thompson_Yb_erasure_gate} based on optical excitations.  Microwave control will open the door to additional robust protocols that can leverage well-developed technologies.

In the original LP-gate~\cite{Levine_Pichler_gate}, a sequence of optical pulses with carefully chosen Rabi frequency, detuning, and relative phase are used to implement a $\mathrm{CZ}$ gate based on the phases accumulated by the computational basis states as rotations occur on the Bloch sphere. We focus here on its generalization based on quantum optimal control~\cite{Jandura_Pupillo_Optimal_control,Pagano_strontium_dCRAB}. Not only are these protocols faster and yield higher fidelity, but they can be made robust to some inhomogeneities.  In addition, we can design waveforms beyond the perfect blockade regime, which can lead to further optimization. 

In the case of cesium considered here, the qubit is encoded in the clock states $\ket{0}\equiv \ket{F=4,m_F=0}$ and $\ket{1}\equiv \ket{F=3,m_F=0}$, and a pulse maps $\ket{F=4,m_F=0}\leftrightarrow \ket{F=3,m_F=1}$ to shelve all of the qubit computational states in the $F=3$ manifold  similar to~\cite{Graham_Saffman_mid_circuit}. The  unpopulated  state $\ket{F=4,m_F=0}$ is then designated as the auxiliary state $\ket{a}$, which can be dressed using a Rydberg laser without coupling any population in the computational states.   This results in the creation of a level diagram as shown in Fig.~\ref{fig:Protocol}. A microwave/Raman laser field is used to couple the qubit state $\ket{1}$ to the dressed state $\ket{\tilde{a}}$, with effective Rabi frequency $\tilde{\Omega}_{\mu w}$, detuning $\tilde{\Delta}_{\mu w}$ and phase $\xi_{\mu w}$. The dressed auxiliary state acts as a proxy to the Rydberg state. We set $\Omega_{\mu w}$ to be constant over the duration of the pulse and choose the detuning $\tilde{\Delta}_{\mu w}=0$ with respect to the dressed state $\ket{\tilde{a}}$. We modulate the phase $\xi_{\mu w}(t)$ of the microwave field over a period of time $\tau$ to implement a $\mathrm{CZ}$ gate.

As shown in Fig.~\ref{fig:Protocol}(d), the computational basis states $\ket{k};k\equiv\{00,01,10,11\}$ do not couple to each other during the evolution. Thus we consider only diagonal gates, a process in which all the basis states evolve  as $\ket{k}\rightarrow\ket{\psi_k(t)}$ and the populations return to these initial states at the end of the pulse, $t=\tau$.  The net result is an accumulation of phase, $\ket{k}\rightarrow\ket{\psi_k(\tau)}= e^{i\phi_k}\ket{k}$. If these phases satisfy the condition $\phi_{11}-2\phi_{01}=\pm\pi$, the gate that is implemented is equivalent to a CZ-gate up to local $e^{-i\phi_{01}\sigma_z/2}$ gates on both qubits. We will refer to this family of gates as ``$\mathrm{CZ}$ gate" hereon for simplicity. Hence, for a unitary two-qubit gate that we implement, $U$, we define the $\mathrm{CZ}$ gate's Bell-state fidelity as 
\begin{widetext}
\begin{equation}
    \mathcal{F}=|1+e^{-i\phi_{01}}\bra{01}U\ket{01}+e^{-i\phi_{10}}\bra{10}U\ket{10} -e^{-i\phi_{11}}\bra{11}U\ket{11}|^2/16.
\end{equation}
\end{widetext}
Note that $\phi_{00}=0$ and $\phi_{01}=\phi_{10}$.  Also note that because all the population begins and ends in the $F=3$ hyperfine manifold, the dressing lasers can be turned on and off rapidly, without the need for adiabatic ramping.

To implement a $\mathrm{CZ}$ gate, we design a control waveform to modulate the microwave phase $\xi_{\mu w}(t)$.  As a proof-of-principle, we choose the waveform to be $N$ piece-wise constant pulses  such that $\xi_{\mu w}(t)=\xi_i$ if $t\in [i\frac{\tau}{N},(i+1)\frac{\tau}{N})$. This waveform implements a unitary $U(\vec{\xi})=\prod_i U(\xi_i)$. Our objective is to find a $\vec{\xi}$ such that $U[\vec{\xi}]$ is locally equivalent CZ-gate unitary. We cast this problem as a minimization problem where we are trying to minimize the cost function $-\mathcal{F}[\vec{\xi}]$. The ``Gradient Ascent Pulse Engineering" (GRAPE) algorithm helps us efficiently calculate the gradient $-\nabla_{\vec{\xi}}\mathcal{F}$ which greatly speeds up minimization algorithms~\cite{Khaneja_Glaser_2005_GRAPE}. Using GRAPE, we can find a waveform that implements a CZ-gate with arbitrarily close to unit fidelity as long as the gate time $\tau$ is larger than the so-called ``quantum speed limit" (QSL), denoted by $\tau_*$.  We seek to choose parameters such that $\tau_*$ is as small as possible, given physical constraints.  We will show that this is achieved by going beyond the regime of a perfect spin-flip blockade, and choosing $J\approx\Omega_{\mu w}$.

\subsection{Optimal dressing strength}\label{sec:optimal_dressing}

\begin{figure*}
    \makebox[\textwidth]{
    \includegraphics[width=\textwidth]{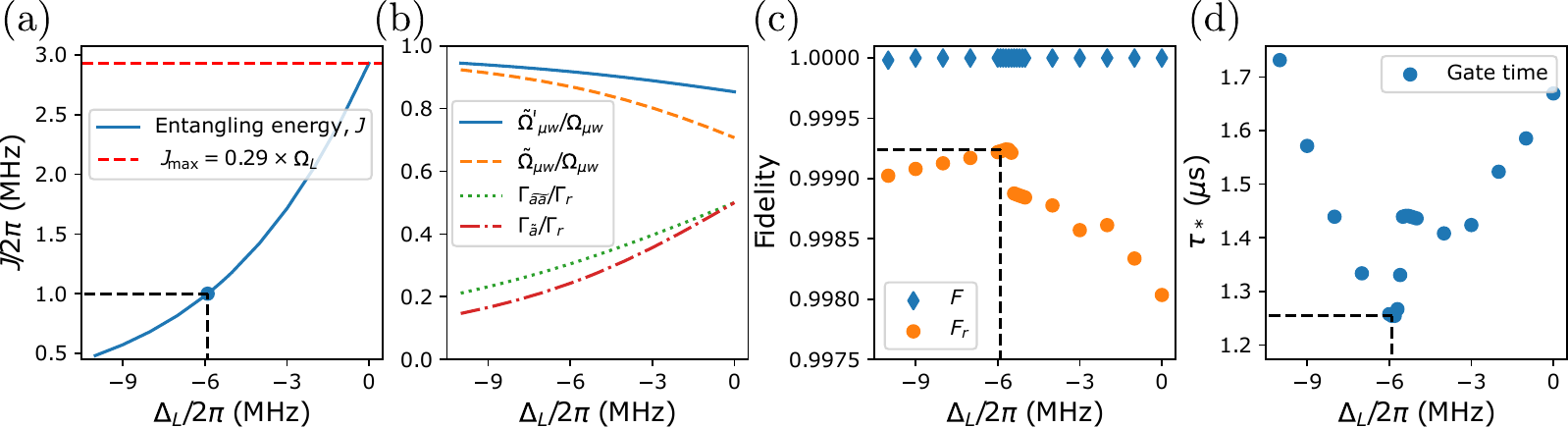}
    }
    \caption{ The effects of the Rydberg-laser detuning on the entangling gate implemented for $\Omega_{\mathrm{L}}/2\pi=10\,\mathrm{MHz}$, $\Omega_{\mu w}/2\pi=1\,\mathrm{MHz}$. (a)  The behavior of $J$ as a function of Rydberg detuning $\Delta_\mathrm{L}$. $J_{\mathrm{max}}=0.29\Omega_\mathrm{L}$ at resonant dressing and falls rapidly with the magnitude of the detuning. Around $\Delta_\mathrm{L}/2\pi=-5.9\,\mathrm{MHz}$, we satisfy the condition $J\approx\Omega_{\mu w}$. (b) As the magnitude of laser detuning is increased, the effective Rabi frequencies, $\tilde{\Omega}_{\mu w}$ and $\tilde{\Omega}'_{\mu w}$ increase and the  effective decay rates, $\Gamma_{\tilde{a}}$ and $\Gamma_{\widetilde{aa}}$, decrease due to decreased admixture of the Rydberg state in the dressed states $\ket{\tilde{a}},\ket{\widetilde{aa}}.$  (c) Using optimal control, a CZ gate waveform can be found at all laser detunings, as shown by the unit theoretical gate fidelity orange dots. The Rydberg decay limited fidelity is shown in blue diamonds, with the highest occurring around the point where $J\approx\Omega_{\mu w}$, at $\mathcal{F}_r=0.9992$ for $1/\Gamma_r=150\,\mu s$. (d) The gate time is also shortest around $J\approx\Omega_{\mu w}$ where $\tau\approx1.26\,\mu \mathrm{s}$. This speedup is due to a combination of larger effective Rabi frequencies compared to resonant dressing and the moderate blockade speedup effect shown in \cref{appendix_1}.
    }
    \label{fig:Optimal_dressing}
\end{figure*}

In the absence of errors, one can reach arbitrary fidelity using optimal control.  The fundamental source of error  is decoherence due to the finite lifetime on the Rydberg state  $1/\Gamma_r$ which is of the order of $100 \mu s$ for typical principal quantum numbers $n\sim 50$ including decay stimulated by blackbody photons.  Because the branching ratio for the decay of population back to the computational subspace is small, we can approximate the master equation simply by a non-Hermitian effective Hamiltonian $H_\mathrm{eff} = H-i\frac{\Gamma_r}{2}\sum\ket{r}\bra{r}$, giving a decoherence limited fidelity $\mathcal{F}_r$.  Furthermore, we define the time spent in the Rydberg state as $T_r=(T_{01}+T_{10}+T_{11})/4$, where $T_{k}=\int_0^\tau\sum_{i=1,2}|\bra{r_i}\ket{\psi_k(t)}|^2dt$ and $\ket{r_1},\ket{r_2}$ are the Rydberg states of the two atoms. We shall call $T_r$ Rydberg time hereon for brevity. These are related as $\mathcal{F}_r\approx 1-\Gamma_r T_r$.  $T_r$ and $\mathcal{F}_r$ thus act as figures of merit with which to compare various protocols.

In the protocol above, the detuning of the Rydberg laser affects the strength of the entangling energy, $J$, as well as the decay rates $\Gamma_\mathrm{\tilde{a}},\Gamma_{\widetilde{aa}}$ and the effective Rabi frequencies $\tilde{\Omega}_{\mu w}, \tilde{\Omega}_{\mu w}'$. In the context of adiabatic dressing gates, we have shown that the optimal operating point was to tune the Rydberg laser toward resonance.  Far off resonance, the rapid decrease in $|J|$ outweighed the decrease in decoherence~\cite{Mitra_Martin_gate}.  For this diabatic optimal control protocol based on the spin-flip blockade, there are additional considerations: the cost of a smaller entangling energy for larger detunings can be offset by larger $\tilde{\Omega}_{\mu w}, \tilde{\Omega}_{\mu w}'$ and smaller $\Gamma_{\tilde{a}}, \Gamma_{\widetilde{aa}}$. Below, we will show that if $J_{\mathrm{max}}>\Omega_{\mu w}$, there is an optimal point and when the dressing is such that $J\sim\Omega_{\mu w}$ we get the minimum $T_r$. The waveforms that minimize $T_r$ also generally minimize the total gate time $\tau_*$. If $J_{\mathrm{max}}<\Omega_{\mu w}$, which is the experimentally less likely scenario, resonant dressing remains optimal.

We demonstrate this optimal dressing for an example parameter set: $\Omega_\mathrm{L}/2\pi = 10$ MHz, $\Omega_{\mu w}/2\pi = 1$ MHz and a perfect Rydberg blockade, $V_{rr}\approx \infty$. For resonant Rydberg dressing at $\Delta_\mathrm{L}=0$, the value of $|J|$ peaks at $J_{\mathrm{max}}=(2-\sqrt{2})\Omega_\mathrm{L}/2\approx 0.29 \Omega_\mathrm{L}$~\cite{Jau_Nature_2016} (see Fig. \ref{fig:Optimal_dressing}(a)) and the dressed state $\ket{\tilde{a}}=(\ket{a}+\ket{r})\sqrt{2}$ has the strongest Rydberg character leading to the weakest effective microwave Rabi frequencies, $\tilde{\Omega}_{\mu w}=\Omega_{\mu w}/\sqrt{2},\tilde{\Omega}'_{\mu w}=\frac{1}{2}(1+\frac{1}{\sqrt{2}})\Omega_{\mu w}$, and largest effective decay rates $\Gamma_{\tilde{a}}=\Gamma_{\widetilde{aa}}=\Gamma_r/2$. As we increase the laser detuning, reducing the dressing strength, we reduce the value of $J$ along with the Rydberg character in the dressed state $|\langle r |\tilde{a}\rangle|^2$, thus reducing the photon scattering rate and increasing $\tilde{\Omega}_{\mu w}$. At various dressing strengths we plot the effective Rabi frequencies $\tilde{\Omega}_{\mu w},\tilde{\Omega}'_{\mu w}$, and decay rates $\Gamma_{\tilde{a}},\Gamma_{\widetilde{aa}}$ in Fig. \ref{fig:Optimal_dressing}(b). 

\begin{figure*}
  \makebox[\textwidth]{
    \includegraphics[width=1\textwidth]{ 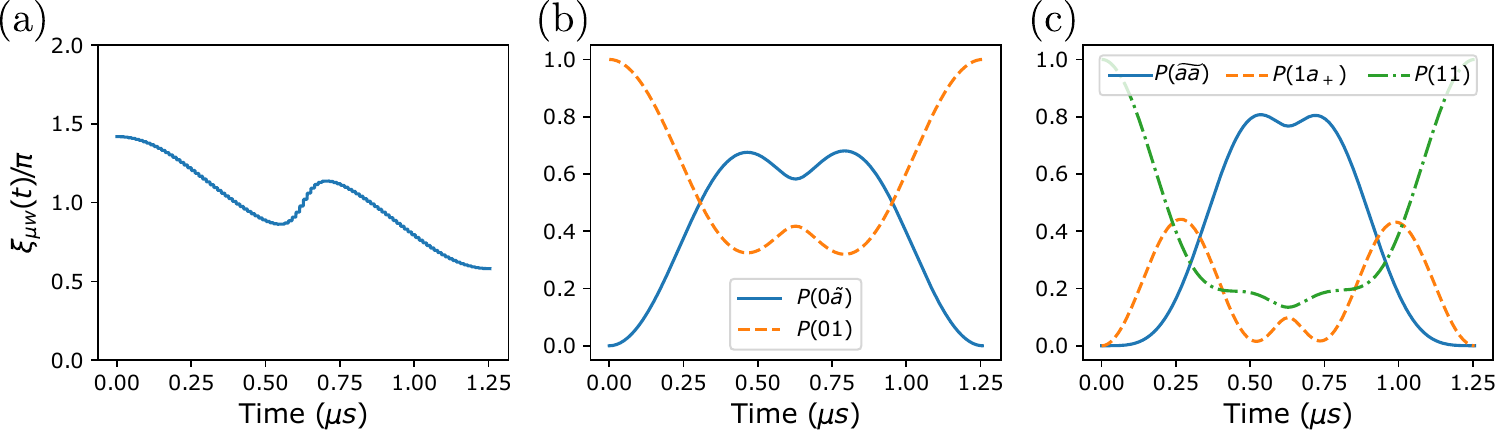}
  }
  \caption{Simulation of a CZ-gate implemented through microwave phase modulation driving Rydberg-dressed hyperfine states. Parameters:  $\Omega_\mathrm{L}/2\pi = 10\,\mathrm{MHz}, \Delta_\mathrm{L}/2\pi= -5.9\,\mathrm{MHz},  \Omega_{\mu w}/2\pi = 1\,\mathrm{MHz}$. For these dressing parameters, $J= 2\pi\times 0.999\,\mathrm{MHz}\approx\Omega_{\mu w}$, corresponding to the optimal point in Fig.~\ref{fig:Optimal_dressing}.
  (a) The phase waveform of the microwave field, $\xi_{\mu w}(t)$ found using GRAPE optimization.
  (b) The populations in various levels during the evolution of state  when initially in the logical $\ket{01}$ state (also the same for $\ket{10}$) when only one atom can be excited by the microwave. 
  (c) The population occupations during the evolution of state when initially in the logical $\ket{11}$ state and two atoms can be excited by the microwave. We work here beyond the regime of the perfect spin-flip blockade.  In the middle portion of the gate, most of the population gets pumped into $\ket{\widetilde{aa}}$ where it accumulates the entangling phase  due to the nonlinear light shift, $J$.
  }
    \label{fig:Waveform}
\end{figure*}

We use GRAPE to find the QSL for each of these scenarios and plot the QSL gate time $\tau_*$ and decoherence limited fidelity $\mathcal{F}_r$ in Fig. \ref{fig:Optimal_dressing}(c,d). We can see the best operating point at $\Delta_\mathrm{L}/2\pi =-5.9\,\mathrm{MHz}$, where $J=\Omega_{\mu w}=2\pi \times 1\,\mathrm{MHz}$. This corresponds to a blockade ratio of $J/\Omega_{\mu w}= 1$. We find the gate time in this scenario to be $\tau_*=1.3\mu s=1.3 \times 2\pi/\Omega_{\mu w}$, which is approximately $25\%$ faster than the resonant dressing scenario. The phase waveform $\xi_{\mu w}(t)$ of this gate shown in Fig.~\ref{fig:Waveform}(a) is similar to that found in~\cite{Jandura_Pupillo_Optimal_control} for optical control. The gate properties are given in Fig.~\ref{fig:Waveform}(b,c), which shows plots of the population in the various levels as a function of time.  For the $\ket{11}$ state, during the central part of the of waveform, a majority of the population is held in the doubly-excited dressed state $\ket{\widetilde{aa}}$.  This leads to faster gates.

For these parameters, the dressed states are given by $\ket{\tilde{a}}=0.85\ket{a}+0.52\ket{r}$ and $\ket{\widetilde{aa}}=0.81\ket{aa}+0.59\ket{ar}_+$. This results in the effective Rabi rates $\tilde{\Omega}_{\mu w}/2\pi \approx 0.85\,\mathrm{MHz}$, $\tilde{\Omega}'_{\mu w}/2\pi\approx 0.9\,\mathrm{MHz}$. The effective decay rates in this case are $\Gamma_{\tilde{a}}=0.27\Gamma_r$, $\Gamma_{\widetilde{aa}}=0.35\Gamma_r$. For a Rydberg lifetime of $150\,\mu s$, we get a decay-limited-fidelity of $99.92\%$. This is the best $\mathcal{F}_r$ achievable in this dressing-based protocol with our chosen Rabi frequencies as shown in Fig. \ref{fig:Optimal_dressing}(c), due to a combination of fast gate time, moderately large effective Rabi frequencies and moderately small dressed decay rates. 
 
This particular speedup in the moderate blockade regime can be understood from the foundations of Lie-algebraic control theory.  Control in the Hilbert space occurs due to the mixing of noncommuting Hamiltonians whose effect on the dynamics is seen in the Magnus expansion through nested commutators~\cite{Merkel_Deutch_2008_PRA_Optimal_control}.  Here the two noncommuting Hamiltonians are the microwave-driven Rabi oscillations and the nonlinear light shift.  The speed limit is fastest when these effects are on the same order, $J \approx\Omega_{\mu w}$.  A similar effect of approximately $8\%$ gate-time reduction is seen in the optical regime, when $V_{rr} \approx \Omega_\mathrm{L}$   for the optical gates (see \cref{appendix_1}), but $V_{rr}$ is not easily controlled and the doubly excited Rydberg state often leads to other inelastic processes. In contrast, $J$ can be accurately controlled by tuning the Rydberg laser detuning, and has a soft-core dependence on distance (see Fig.~\ref{fig:J_vs_r})~\cite{Johnson_Rolston_dressing,Jau_Nature_2016}.
 
\begin{figure*}
  \makebox[\textwidth]{
    \includegraphics[width=\textwidth]{  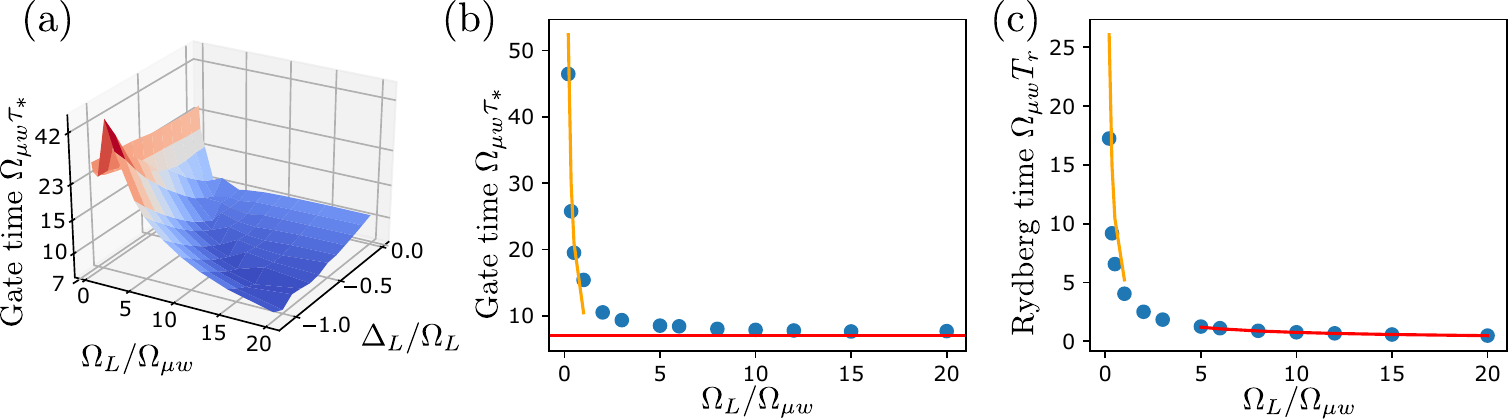}
  }
  \caption{Behavior of gate-time $\tau_*$ and Rydberg time $T_r$ as a function of dressing. (a) Gate time  as a function of $\Omega_\mathrm{L}/\Omega_{\mu w}$ and $\Delta_\mathrm{L}/\Omega_\mathrm{L}$. The narrow valley in the surface plot corresponds to the optimal dressing strengths for a given $\Omega_\mathrm{L}/\Omega_{\mu w}$. For large $\Omega_{\mathrm{L}}/\Omega_{\mu w}$, this valley closely tracks the curve for $\Delta_\mathrm{L}/\Omega_\mathrm{L}$ that results in $J=\Omega_{\mu w}$. In the regime where $.29 \Omega_L <  \Omega_{\mu w}$, the optimal dressing is on laser resonance where $J_{\mathrm{max}}=0.29\Omega_\mathrm{L}$.  (b) Gate times $\tau_*$ as a function of $\Omega_{\mathrm{L}}/\Omega_{\mu w}$. The red (horizontal) line corresponds to the weak-dressing-regime bound, $1.11\times 2\pi/\Omega_{\mu w}$, and orange (vertical) line corresponds to strong-dressing-regime bound, $\pi/J_{\mathrm{max}}$ (c) Rydberg times $T_r$ as a function of $\Omega_{\mathrm{L}}/\Omega_{\mu w}$. The red (horizontal) line corresponds to the weak-dressing-regime bound, $3.5/\Omega_{\mathrm{L}}^{2/3}\Omega_{\mu w}^{1/3}$~(Eq. \ref{eqn:analytical_wdr}), and the orange (vertical) line corresponds to the strong-dressing-regime bound, $1.66\pi/\Omega_L$~(Eq. \ref{eqn:analytical_sdr}).
    }
    \label{fig:2d_plot}
\end{figure*}

More generally, Fig.~\ref{fig:2d_plot}(a) shows a plot of the QSL as a function of the ratio of the laser and microwave Rabi frequencies $\Omega_\mathrm{L}/\Omega_{\mu w}$  and normalized laser detuning, $\Delta_\mathrm{L}/\Omega_\mathrm{L}$. The valley running across the plot corresponds to the optimal dressing strength that minimizes the gate time. This valley reliably corresponds to dressing strengths at which $J=\Omega_{\mu w}$ in the regime where $J_{\mathrm{max}}>\Omega_{\mu w}$. If $\Omega_\mathrm{L}<3.5\Omega_{\mu w}\implies J_{\mathrm{max}}<\Omega_{\mu w}$, we can never achieve $J \sim \Omega_{\mu w}$. In this scenario, resonant dressing is the optimal condition, similar to \cite{Mitra_Martin_gate}, as seen in Fig. \ref{fig:2d_plot}(a). We look at the properties of these optimally dressed gates for different limits of $\Omega_\mathrm{L}/\Omega_{\mu w}$.

 In a realistic experimental scenario, we do not have access to unlimited field strengths. For $^{131}\mathrm{Cs}$ under consideration, we are typically limited by the microwave/Raman Rabi frequency, ($\Omega_{\mu w}<J_{\mathrm{max}}\approx 0.3 \Omega_\mathrm{L}$). For the limiting case $\Omega_\mathrm{L}\gg\Omega_{\mu w}$, the optimal dressing strength is achieved by detuning the laser such that $J\approx\Omega_{\mu w}$. In this case, $\Delta_\mathrm{L} \gg \Omega_\mathrm{L}$,  the weak dressing regime (wdr) where\cite{Jau_Nature_2016}
\begin{equation}
    J^{wdr} \approx \frac{\Omega_\mathrm{L}^4}{8\Delta_\mathrm{L}^3}.
\end{equation}
Setting $ J^{\mathrm{wdr}} =\Omega_{\mu w}$, for weak dressing we find the optimal dressing detuning, 
\begin{equation}
    \Delta^{\mathrm{wdr}}_L=\Omega_\mathrm{L}^{4/3}/2\Omega_{\mu w}^{1/3}
\end{equation} 
Also in this regime, $\tilde{\Omega}_{\mu w}, \tilde{\Omega}_{\mu w}'\rightarrow\Omega_{\mu w}$ (up to first order in $\Omega_\mathrm{L}/\Delta_\mathrm{L}$) and $\Gamma_{\tilde{a}}\rightarrow\frac{\Omega_\mathrm{L}^2}{4\Delta_\mathrm{L}^2}\Gamma_r,\Gamma_{\widetilde{aa}}\rightarrow\frac{\Omega_\mathrm{L}^2}{2\Delta_\mathrm{L}^2}\Gamma_r$. By finding the CZ gate waveform for this scenario, we find that 
\begin{equation}\label{eqn:analytical_wdr}
    T_r^{\mathrm{wdr}} =\frac{7\Gamma_{\tilde{a}}}{\Omega_{\mu w}\Gamma_r}=\frac{3.5}{\Omega_\mathrm{L}^{2/3}\Omega_{\mu w}^{1/3}}.
\end{equation}

When we are not limited by microwave Rabi frequency ($\Omega_{\mu w}>J_\mathrm{max}$),  and $V_{rr}\gg\Omega_{\rm L}$, the best regime of operation is on Rydberg resonance ($\Delta_\mathrm{L}=0$), the strong dressing regime (sdr) similar to \cite{Mitra_Martin_gate,Schine_Kaufman_Bell_state_Martin}.
The gate time $\tau$ is also minimized for this scenario. For this resonant dressing, $\tilde{\Omega}_{\mu w}= \Omega_{\mu w}/\sqrt{2}, \tilde{\Omega}_{\mu w}'=\Omega_{\mu w}(1+1/\sqrt{2})$ and $\Gamma_{\tilde{a}}=\Gamma_{\widetilde{aa}}=\frac{\Gamma_r}{2}$. We consider the limit of very large $\Omega_{\mu w}$, i.e., $\Omega_{\mu w}\gg\Omega_\mathrm{L}$, for which the gate time asymptotically reaches $\pi/J_{\mathrm{max}}$. Hence, we find

\begin{equation}\label{eqn:analytical_sdr}
    T_r^{\mathrm{sdr}}\sim \frac{\pi}{J_{\mathrm{max}}}\frac{(2\Gamma_{\tilde{a}}+\Gamma_{\widetilde{aa}})}{4\Gamma_r}\approx 1.66 \times \frac{\pi}{\Omega_\mathrm{L}}.
\end{equation}
Hence, in the sdr, in our protocol the time spent in the Rydberg state is comparable to that in optical protocols \cite{Jandura_Pupillo_Optimal_control,Pagano_strontium_dCRAB}. In the wdr, the time spent in the Rydberg state $\sim \Omega_\mathrm{L}^{-2/3}\Omega_{\mu w}^{-1/3}$ is generally longer than that achievable by an optically driven gate at the same UV Rabi frequency ($\sim \Omega_\mathrm{L}^{-1}$).

For the Rabi frequencies we considered ($\Omega_\mathrm{L}/2\pi = 10$ MHz, $\Omega_{\mu w}/2\pi = 1$ MHz), the optimal operating regime is moderate dressing, $\Delta_{\mathrm{L}}=-0.59\Omega_{\mathrm{L}}$ (see Fig.~\ref{fig:Optimal_dressing}). For these parameters, the microwave-based gate (optical gate) has a gate time of 1.3 $\mu$s (0.12 $\mu$s) and a Rydberg decay-limited-fidelity of $99.92\%$ ($99.97\%$). While the microwave-based gate has a slightly lower fundamental fidelity, the current state-of-the-art gate fidelities are far from the limit set by the Rydberg lifetime, owing to experimental noise and inhomogeneties. 
The main potential benefit of the microwave-based protocol is the potential to mitigate these with robust optimal control as we demonstrate in the following section.

\section{Robust control}
\label{sec:Robust_control}

The fidelity of entangling gates is fundamentally limited by the finite lifetime of the Rydberg state.  Other important sources of errors include inhomogeneities in the atomic positions and momenta which affect their coupling to external fields and the atom-atom interactions, and high sensitivity of Rydberg states to external electric fields. We categorize these deleterious effects as uncertainty in the Rabi frequencies, and detuning errors, which include Doppler shifts and spurious lightshifts on the Rydberg states. We consider the effects of such noises on our gate protocol, and using the tools of robust control we can further reduce the residual effects of inhomogeneities.  

To study this we include uncertain parameters in the dynamics with the following Hamiltonian associated with two atoms in the $\ket{11}$ state state~\cite{Keating_gate},

\begin{align}
    H&=H_++H_-.\\
    H_+ &= -\Delta_\mathrm{L}^+ (\ketbra{B}{B}+\ketbra{D}{D}) -(\Delta_{\mathrm{L}1}+\Delta_{\mathrm{L}2}-V_{rr})\ketbra{rr}{rr} \nonumber \\
    &+\frac{\Omega^+_\mathrm{L}}{2}(\ketbra{11}{B}+\ketbra{B}{rr}+\mathrm{h.c.}). \\
    H_{-}&= \frac{\Omega^-_\mathrm{L}}{2}(\ketbra{11}{D}+\ketbra{D}{rr}+\mathrm{h.c.})\nonumber\\
    &-\Delta_\mathrm{L}^-(\ketbra{B}{D}+\ketbra{D}{B}) . 
\end{align}
Here $H_+$ is the symmetric part of the Hamiltonian, where the computational state $\ket{11}$ is coupled to the singly-excited  ``bright state,"
\begin{equation}
\ket{B} = (\ket{ra}+\ket{ar})/\sqrt{2} 
\end{equation}
with Rabi frequency 
\begin{equation}
\Omega_\mathrm{L}^+ \equiv (\Omega_{\mathrm{L}1}+\Omega_{\mathrm{L}2})/\sqrt{2}, 
\end{equation}
where $\Omega_{\mathrm{L}1},\Omega_{\mathrm{L}2}$ are the Rydberg Rabi frequencies of the two atoms. Similarly, the bright and dark states are detuned by 
\begin{equation}
\Delta_\mathrm{L}^+ \equiv (\Delta_{\mathrm{L}1}+\Delta_{\mathrm{L}2})/2, 
\end{equation}
where $\Delta_{\mathrm{L}1},\Delta_{\mathrm{L}2}$ are the effective Rydberg detunings of the two atoms. We now look at $H_-$, the asymmetric part of the Hamiltonian. Due to unequal Rabi frequencies, the so-called ``dark state" participates in the dynamics
\begin{equation}
\ket{D} = (\ket{ra}-\ket{ar})/\sqrt{2}.
\end{equation}
which couples to $\ket{11}$ with Rabi frequency

\begin{equation}
\Omega_\mathrm{L}^- \equiv (\Omega_{\mathrm{L}1}-\Omega_{\mathrm{L}2})/\sqrt{2}.
\end{equation}  

In addition, the bright and dark states are coupled to one another due to the relative detuning difference ($\Delta_L^-=(\Delta_{\mathrm{L}1}-\Delta_{\mathrm{L}2})/2$) as this affects the relative phase of the laser seen by the two atoms. 
We note that the single-atom excitations driven from the logical states  $\ket{01}$ and  $\ket{10}$ will also have the modified Rydberg detuning and laser Rabi frequency.

\begin{figure*}
  \makebox[\textwidth]{
    \includegraphics[width=\textwidth]
    {  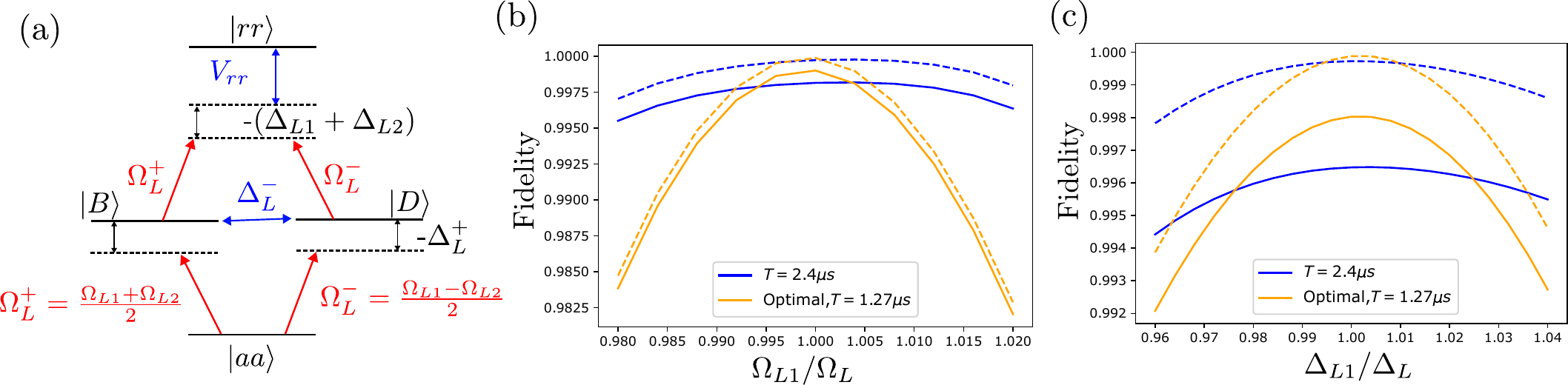}}
  \caption{Effects of Rydberg inhomogeneities and mitigating them via robust control methods. (a) Level diagram for two-atom dressing in the presence of imperfect Rabi frequencies and detuning errors. The asymmetric dressing of the two atoms results in unequal Rydberg-laser Rabi frequencies, $\Omega_{\mathrm{L}1},\Omega_{\mathrm{L}2}$, and coupling from $\ket{aa}$ to both the bright state $|B\rangle = |ar\rangle_+$ and the dark state $
    |D\rangle = |ar\rangle_-$,  with Rabi frequencies $\Omega_\mathrm{L}^+=\frac{\Omega_{\mathrm{L}1}+\Omega_{\mathrm{L}2}}{\sqrt{2}}$ and $\Omega_\mathrm{L}^-=\frac{\Omega_{\mathrm{L}1}-\Omega_{\mathrm{L}2}}{\sqrt{2}}$ respectively. Similarly, the atoms see unequal Rydberg detunings $\Delta_{\mathrm{L}1},\Delta_{\mathrm{L}2}$ due to a combination of Doppler noise, spurious Rydberg light shifts etc. The Bright and Dark states are effectively detuned by $\Delta_\mathrm{L}^+=(\Delta_{\mathrm{L}1}+\Delta_{\mathrm{L}2})/2$, and coupled by a Rabi frequency $\Delta_\mathrm{L}^-=(\Delta_{\mathrm{L}1}-\Delta_{\mathrm{L}2})/2$. Because the dressed state $\ket{\widetilde{aa}}$ is an eigenstate of the total Hamiltonian in this subspace, the net effect of these noises together is uncertainty in the light-shifts and the dressed microwave-Rabi frequencies.  The dominant effect is the uncertainty in the entangling energy, $J$. (b) Sensitivity to uncertainty in $\Omega_{\mathrm{L}1}$ for the optimal gate (orange steeper parabola, $T=1.27 \mu s$) shown in Fig.~\ref{fig:Waveform} and the robust gate (blue flatter parabola, $T=2.4 \mu s$) found by using robust optimal control. The dashed lines correspond to the ideal fidelity as function of inhomogeneity, assuming unitary evolution, while the solid lines include decoherence to to Rydberg decay. We choose  $\Omega_{\mathrm{L}1}/\Omega_L=\Omega_{\mathrm{L}2}/\Omega_{\mathrm{L}}$ ($\Omega_{\mathrm{L}}$ is the ideal Rabi frequency). The robust waveform achieves significant resilience to uncertainty in $\Omega_{\mathrm{L}}$ at the price of longer gate times.  This implies a trade-off with Rydberg decay.  The regime in which robust control is useful is seen for large inhomogeneities, where the solid blue curve is above the orange curve. (c) Sensitivity to uncertainty in $\Delta_{\mathrm{L}1}$, with  $\Delta_{\mathrm{L}2}=0$. The fidelity as a function of $\Delta_{\mathrm{L}1}/\Delta_{\mathrm{L}}$ ($\Delta_{\mathrm{L}}$ is the ideal Rydberg detuning) using the same waveform that optimizes robustness against inhomogeneity in $\Omega_{\mathrm{L}1}$ shows significant robustness to detuning noise. Typically, a gate optimized to be robust against one noise source has increased sensitivity against other noises. In this case however, since both noises are effectively the same uncertainty in the dressed state $\ket{\widetilde{aa}}$, we are able to achieve robustness against both major noises.
    }    
    \label{fig:Robust_control}
\end{figure*}
 
The net effect of these noise sources on dressing and
the microwave-driven gates is as follows. Firstly, Doppler shifts at microwave frequencies are negligible due to a much smaller microwave wavenumber. 
The major source of Doppler noise is the dressing laser itself, due to a wavenumber $k_L = 2\pi/\lambda_\mathrm{L}$ which $10^5$ times larger. 
This affects the system in two ways:  (1) the dressed states themselves are modified; (2) the light shifts and associated entangling energies are modified.  
The dressed states $\ket{0\tilde{a}}$, $\ket{\tilde{a} 0}$, and $\ket{1\tilde{a}}_+$ receive a one-atom light shift, only slightly modified due to the Doppler shift when the excitation to the Rydberg state is sufficiently far from resonance, $\Delta_\mathrm{L}\gg P_\mathrm{com}k_L/(2m)$. 
The dressed state $\ket{\widetilde{aa}}$ receives a two-atom light shift.  
This arises as  $\ket{aa}$ is coupled to the bright and dark states, $\ket{B}$ and $\ket{D}$, which are themselves coupled together through the relative motion as shown in Fig.~\ref{fig:Robust_control}(a); here we assume a perfect Rydberg blockade. 
This leads to a slightly different dressed state $\ket{\widetilde{aa}}$ as compared to the ideal case, which is now an admixture of $\ket{aa}$, $\ket{B}$, and $\ket{D}$ with amplitudes that depend on the random motion of the particles. 
The strength of the entangling interaction $J$ is modified as well.  
The changes in the dressed state alters the dressed Rabi frequencies $\tilde{\Omega}_{\mu w},\tilde{\Omega}_{\mu w}'$ but this change is much weaker than the change in $J$, as seen in Fig.~\ref{fig:Optimal_dressing}. 
The dominant noise effect due to thermal motion is thus the uncertainty in the value of $J$.  

Given how the dressed states are modified by inhomogeneities, we can determine their effect on gate fidelity.  As discussed in Sec. II, there are two possible regimes of operation.   Under the condition of a strong spin-flip blockade, when $J\gg\tilde{\Omega}'_{\mu w}$, the actual value of $J$ is unimportant, and its small uncertainty is negligible.  The residual effect of inhomogeneities is on the one-atom light shifts.  This, again, is negligible when the laser detuning is large compared to the Doppler width.  In this regime, the microwave-drive dressed gates are intrinsically robust to small thermal fluctuations.   Note, however, that this is not necessarily the regime of the highest fidelity due to the finite Rydberg lifetime. In Sec. II we found that the fastest gates with the smallest integrated Rydberg time were achieved when $J\approx \tilde{\Omega}'_{\mu w}$.  In that regime, we must consider the uncertainty in $J$.  To do so we can employ the methods of ``robust optimal control."

We consider the method of inhomogeneous control originally introduced by Khaneja~\cite{Anderson_Jessen_Robust_control,Khaneja_inhomogeneous_PRA_2006}. The key idea is to optimize the average fidelity $\bar{\mathcal{F}}=-\int d\vec{\lambda} P(\vec{\lambda})\mathcal{F}(\vec{\lambda})$, where $\mathcal{F}(\vec{\lambda})$ is the fidelity at a fixed set of the uncertain parameters, and $P(\vec{\lambda})$ is their probability distribution. We again use GRAPE to optimize the modified cost function $\bar{\mathcal{F}}=[2\mathcal{F}(\Omega_1=\Omega_2=\Omega_\mathrm{L})+\mathcal{F}(\Omega_1=\Omega_2=1.02\Omega_\mathrm{L})+\mathcal{F}(\Omega_1=\Omega_2=0.98\Omega_\mathrm{L})]/4$ where $\Omega_\mathrm{L}$ is the ideal Rydberg Rabi frequency. We are effectively optimizing the fidelity in a $\pm2\%$ window of the ideal Rabi frequency, but with a greater weight at the ideal parameters. We expect to get better robustness as we increase the gate time, but this generally worsens the sensitivity against other noise. In optical control protocols, waveforms that are robust against both Rabi frequency uncertainty and Doppler noise require significantly longer gate times, as shown in~\cite{jandura_Pupillo_robust, fromonteil_Bluvstein_Pichler_robust,Mohan_Kokkelmans_Robust_control_PRR_2023}. In the dressed-state case we find that pulses that are robust against Rabi frequency noise are also robust against Doppler detuning noise, as shown in Fig.~\ref{fig:Robust_control}(b,c). We attribute this to the fact that effectively, both noises manifest as uncertainty in $J$ and we are actually engineering robustness against this one quantity which manifests as reduced sensitivity to uncertainty in both $\Omega_\mathrm{L}$ and $\Delta_\mathrm{L}$.

\section{Summary and Outlook}
\label{sec:summary_and_outlook}
We proposed a new method to implement entangling gates between neutral atoms based on the spin-flip blockade.  This has the potential advantage that all coherent control can be performed robustly with microwave fields.  We showed for the specific case of alkali atoms, such as cesium, by optically dressing an auxiliary hyperfine state with Rydberg character, the level structure associated with the optically-excited Rydberg blockade can be mapped to the microwave-excited spin-flip blockade.  In this way any protocol that could be implemented with optical control can be mapped into the microwave regime.

We studied the specific example of the spin-flip version of the generalized Levine-Pichler gate, implemented through modulation of the microwave phase, whose waveform is found by quantum optimal control.   In contrast to the optical regime,  where the doubly-excited Rydberg potential $V_{rr}$ is not well defined at a short interatomic distances, in the microwave regime we consider going beyond the perfect blockade regime.  This is profitable as the doubly spin-excited state is well defined, and the blockade strength is a much more stable function of interatomic motion due to its soft-core nature.  We find that the optimal quantum speed limit occurs when the entangling energy $J$ is approximately equal to the driving microwave Rabi frequency $\Omega_{\mu w}$.  As $J$ is controllable by laser detuning, we find that for a laser Rabi frequency $\Omega_{L} \gg \Omega_{\mu w}$, we can achieve high fidelity in the off-resonance regime.  For $\Omega_{\mu w}/2\pi = 1$ MHz, we can achieve a fidelity $>99.9\%$, limited solely by the finite lifetime of the Rydberg state.  While higher fidelity is in principle possible with optically controlled gates, in practice, the fidelity is far from the theoretical maximum, limited otherwise by technical noise sources.  The robustness of microwave control may allow us to reach closer to the fundamental limits.

Of particular importance are the noises arising from atomic thermal motion which gives rise to Doppler shifts, spurious light-shifts on Rydberg states and imperfect addressing of atoms. 
The overall effect is that laser detuning and amplitude are then uncertain.  The dominant net effect of this noise boils down to uncertainty in one parameter, the entangling energy, $J$. We show that this allows us to use the method of inhomogeneous control to engineer robustness against both laser amplitude and detuning noise simultaneously, with only a modest increase in gate time. 

These conclusions extend to other forms of noise such as fluctuations in laser intensity (Rabi frequency), and phase (detuning). Together, these noises leads to inhomogeneous broadening on the timescale $T_2^*$.  When $T_2^* \ll 1/\Gamma_r$,  off-resonance dressing, $\Delta_L \gg 1/T_2^*$, enables higher fidelity gates so long as there is sufficient laser power that we are not too far off resonance, e.g.,  $\Delta_L \sim \Omega_L$, and $J$ remains large compared the dressed-state decay rate.

While we have focused here on the microwave version of the generalized LP-gate, in principle any protocol developed in the optical regime can be mapped into the microwave regime.  For example, the recently studied protocol involving adiabatic rapid passage beyond the perfect blockade~\cite{Mitra_Omanakuttan_adiabatic_gates} can be mapped to a microwave/Raman adiabatic passage.  Such gates might achieve even higher fidelity as they can take full advantage of the entangling energy $J$.  Overall, the scheme proposed here should enable exploration of different tradeoffs in speed, noise, and robustness in order to obtain the highest fidelity entangling gates.

\section{Acknowledgements}
We thank Matt Chow, Bethany Little, Anupam Mitra, Sven Jandura and Bharath Hebbe Madhusudhana for helpful discussions. This work is supported by Sandia Laboratory Directed Research and Development (LDRD) and Quantum Systems through Entangled Science and Engineering (Q-SEnSE) funded by National Science Foundation Quantum Leap Challenge Institute (NSF QLCI) grant no. 2016244. Sandia National Laboratories is a multimission laboratory managed and operated by National Technology \& Engineering Solutions of Sandia, LLC, a wholly owned subsidiary of Honeywell International Inc., for the U.S. Department of Energy’s National
Nuclear Security Administration under contract DENA0003525. This paper describes objective technical results and analysis. Any subjective views or opinions that might be
expressed in the paper do not necessarily represent the views of the U.S. Department of Energy or the United States Government.

\appendix

\section{} \label{appendix_1}

\begin{figure}[ht!]
    \centering
    \includegraphics[width=\columnwidth]{ 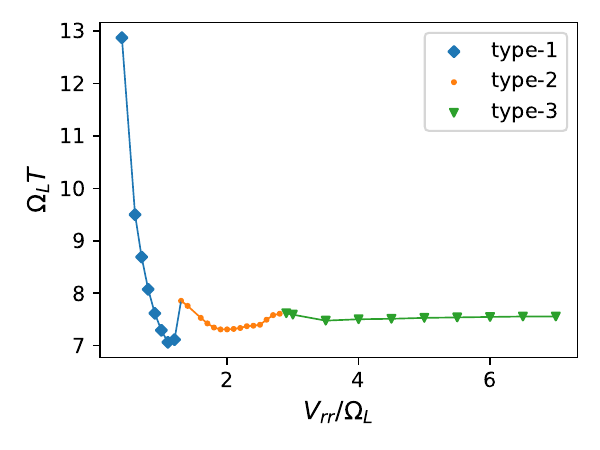}
    \caption{Entangling gate time vs Rydberg blockade strength $V_{rr}/\Omega_\mathrm{L}$ for optical control gates. The waveforms for the gates are calculated using GRAPE. For a strong blockade, the gate time approaches $\tau\Omega_\mathrm{L}=7.6$. For a small blockade ratio, the gate time approaches $\tau\Omega_\mathrm{L}=\pi/V_{rr}$. At the sweet spot of moderate blockade strength $V_{rr}/\Omega_\mathrm{L}\approx1$, the gate time is $7/\Omega_\mathrm{L}$. The three different colors/markers correspond to three different classes of waveforms, differentiated by their shapes\cite{Jandura_private_comm}.}
    \label{fig:QSL}
\end{figure}

In our studies of the microwave-driven gates on dressed states, we have seen that optimal gates can be designed when $J/\Omega_{\mu w} \approx 1$. This maps to $V_{rr}/\Omega_r\approx1$ in the optical regime. While this moderate blockade optical protocol is not experimentally common due to the difficulty in controlling $V_{rr}$, analyzing it further provides valuable insight into our gate protocol. In Fig.~\ref{fig:QSL} we show the QSL gate time as a function of the blockade ratio for the optical protocols at various $V_{rr}/\Omega_\mathrm{L}$. We use GRAPE to find $\xi_L(t)$ laser-phase waveforms that implement $\mathrm{CZ}$ gates similar to \cite{Jandura_Pupillo_Optimal_control,Pagano_strontium_dCRAB} at various $V_{rr}/\Omega_\mathrm{L}$. Perhaps surprisingly, we see that a blockade ratio as low as $V_{rr}/\Omega_\mathrm{L}\approx1$ doesn't make the gate much slower; In fact, the gates are slightly faster at this moderate blockade ratio. At large blockade ratios, the state $\ket{rr}$ is essentially absent from the dynamics which explains the saturation of the gate time at $7.6/\Omega_\mathrm{L}$ and these gates were studied in \cite{Jandura_Pupillo_Optimal_control,Pagano_strontium_dCRAB}. On the other end, for very weak blockade ratios, the entangling energy is simply too small, and a larger gate time is needed to compensate. At very small $V_{rr}/\Omega_\mathrm{L}$, the gates approach a protocol similar to \cite{Keating_gate} where the populations are driven from computational states $\ket{01},\ket{11}$ into the Rydberg states $\ket{0r},\ket{1r}_+$ and kept there until they accumulate the right amount of dynamical phases, $\phi_{11}-2\phi_{01}\approx (E_{rr}-2E_r)\tau=\pi$. The gate time asymptotically approaches $\pi/V_{rr}$ in this limit (similar to the weak-blockade limit in \cite{Mitra_Omanakuttan_adiabatic_gates}). Near the optimal point of $V_{rr}\approx\Omega_\mathrm{L}$, the entirety of the nonlinear phase accumulated, $\phi_{11}-2\phi_{01}$ is geometrical in nature, with no contribution from dynamical phases. This explains the fast gate time, as these geometrical gates tend to follow the geodesic paths in the Hilbert space, resulting in optimal gate times\cite{Bharath_PRL}. This counter-intuitive result explains why choosing a dressing regime where $J\approx\Omega_{\mu w}$ would give us the optimal gate times, which is borne out by the results in Fig.~\ref{fig:Optimal_dressing}(d).

\bibliography{references}

\end{document}